\documentclass [12pt,a4paper]{article}
\usepackage{bbm}
\usepackage{amsfonts}
\usepackage{mathrsfs}
\usepackage {amssymb}
\usepackage {amsmath}
\usepackage{amsthm}
\usepackage{latexsym}
\usepackage {amssymb}
\usepackage {amsmath}
\makeatletter
 \@addtoreset{equation}{section}
 
\makeatother

\usepackage{lastpage}
\usepackage{epsfig}

\begin{document}
\begin{center}
\textbf{\Large{Generalized Hamming Weights of Irreducible Cyclic
Codes} }

\end{center}

\begin{center}
\small Minghui Yang,  Jin Li,  Keqin Feng, Dongdai Lin
\end{center}


\noindent\textbf{Abstract}-The generalized Hamming weight (GHW)
$d_r(C)$ of linear codes $C$ is a natural generalization of the
minimum Hamming distance $d(C)(=d_1(C))$ and has become one of
important research objects in coding theory since Wei's originary
work [23] in 1991. In this paper two general formulas on $d_r(C)$
for irreducible cyclic codes are presented by using Gauss sums and
the weight hierarchy $\{d_1(C), d_2(C), \ldots, d_k(C)\}$ $(k=\dim
C)$ are completely determined for several cases.

\noindent\textbf{keywords}-generalized Hamming weight, irreducible
cyclic code, Gauss sum.

\section{Introduction}

\ \ \ \ Let $\mathbb{F}_q$ be the finite field with $q$ elements. A linear
code $C$ with parameters $[n, k]_q$ means that $C$ is an
$\mathbb{F}_q$-vector subspace of $\mathbb{F}_q^n$ with dimension
$k=\dim_{\mathbb{F}_q}C$. For each $r$, $0\leq r\leq k$, we denote
the set of all $r$-dimensional $\mathbb{F}_q$-subspaces of $C$ by
$\begin{bmatrix}C\\r
\end{bmatrix}_q.$
For each $V\in\begin{bmatrix}C\\r
\end{bmatrix}_q,$ the support of $V$ is defined by
$$\textrm{Supp}(V)=\{i:0\leq i \leq n-1,\textrm{there exists} \ c=(c_0, c_1, \ldots, c_{n-1})\in V \ \textrm{such that} \ c_i\neq 0\}.$$
The $r$-th generalized Hamming weight (GHW) of a $q$-ary linear code $C$ is defined by
$$d_r(C)=\min\left\{|\textrm{Supp}(V)|: V\in\begin{bmatrix}C\\r
\end{bmatrix}_q\right\}$$
and $\{d_1(C), d_2(C), \ldots, d_k(C)\}$ is called the weight
hierarchy of $C(d_0(C)=0).$ It is obvious that $d_1(C)$ is just the
minimun Hamming distance $d(C)$.

The concept on GHW has appeared as early as in 1970's ([12,17]) and
has become an important research object in coding theory after Wei's
paper [23] in 1991, where Wei gives a series of beautiful results on
GHW and indicates that it completely characterizes the performance
of a linear code when it is used on wire-tap channel of type II
which has connection with cryptography. GHW is also used to deal
with $t$-resilient functions and trellis or branch complexity of
linear codes [20].

In past two decades the value of GHWs has been determined or estimated for many
series of linear codes (RM codes [10,23], BCH codes [4,8,9], trace
codes [19], cyclic codes [6,16], AG codes [2,5,7,15,18,25,26],
binary Kasami codes [13] and other codes [22,24]) and the weight
hierarchy is totally determined in several cases. The bounds,
asymptotic behaviour and duality of GHWs have been found
[1,11,17,20,23]. But in general speaking, to determine the weight
hierarchy is a difficult problem.

In this paper we deal with GHWs of $q$-ary irreducible cyclic
codes. For binary case $(q=2)$ there exist several results in
[14,20,21]. We consider the general case where $q$ is any power of a
prime number. We firstly present two general formulas on $d_r(C)$
((3.3) and (3.4) in Theorem 3.2) which involves Gauss sums and
character sum $\sum_{\beta\in H\backslash \{0\}}\varphi(\beta)$,
where $\varphi$ is a multiplicative character of $\mathbb{F}_Q$
$(Q=q^k)$ and $H$ is a $\mathbb{F}_q$-subspace of $\mathbb{F}_Q$
with dimension $r$ or $k-r$. When involved Gauss sums can be
calculated and have the same values, the character sum is reduced to
be the size of $H\bigcap\langle\theta^{e'}\rangle$, where
$\langle\theta^{e'}\rangle$ is a subgroup of
$\mathbb{F}_Q^{\ast}=\langle\theta\rangle$. Then we can determine
$d_r(C)$ for smaller $r$ by (3.3) and larger $r$ by (3.4). And the
weight hierarch $\{d_r(C): 1\leq r \leq k\}$ can be totally
determined in several cases.

The paper is organized as follows. In Section 2 we introduce several basic facts on Gauss sums and previously known results on GHWs. Then we present
two general formulas on $d_r(C)$ and their direct consequences in Section 3. In Section 4 we obtain more results on $d_r(C)$ for several particular
cases. Section 5 is conclusion.

\section{Preliminaries}

\subsection{Gauss sums}

\ \ \ \  We introduce several basic facts on Gauss sums used in this paper. For more details on Gauss sums we refer the book [3].

Let $q=p^l$ where $p$ is a prime number and $l\geq 1,$ $\theta$ be a primitive element of the finite field $\mathbb{F}_q$,
namely $\mathbb{F}_q^{\ast}=\langle\theta\rangle$. Let $\zeta_m=e^{\frac{2\pi\sqrt{-1}}{m}}$ for any positive integer $m$.
The group of additive characters of $\mathbb{F}_q$ is
$$\hat{\mathbb{F}_q}=\{\lambda_b: b\in\mathbb{F}_q\},$$
where $$\lambda_b(x)=\zeta_p^{T(bx)} \ \ \ \ (x\in\mathbb{F}_q)$$
and $T$ is the trace mapping from $\mathbb{F}_q$ to $\mathbb{F}_p$. The group of multiplicative characters of $\mathbb{F}_q$
is
$$  \hat{\mathbb{F}_q^{\ast}}=\{\psi^i: 0 \leq i\leq q-2\}=\langle\psi\rangle  $$
where $\psi$ is defined by $\psi(\theta)=\zeta_{q-1}.$

For each $\chi\in\hat{\mathbb{F}_q^{\ast}}$ and $\lambda\in\hat{\mathbb{F}_q}$ we define the Gauss sum on $\mathbb{F}_q$ by
$$G(\chi, \lambda)=\sum_{x\in\mathbb{F}_q^{\ast}}\chi(x)\lambda(x)\in Z[\zeta_{p(q-1)}].$$

 \textsl {Lemma 2.1:} (1) (trivial cases)
\begin{displaymath}
 G(\chi, \lambda)=G_q(\chi, \lambda)
 = \left\{ \begin{array}{ll}
q-1, & \textrm{if $\chi=1(=\psi^0)$ and $\lambda=1(=\lambda_0)$ }\\
-1, & \textrm{if $\chi=1$ and $\lambda\neq1$ }\\
0, & \textrm{if $\chi\neq1$ and $\lambda=1$. }
\end{array} \right.
\end{displaymath}

(2) For $b\in\mathbb{F}_q^{\ast}$ (namely, $\lambda_b\neq1$)

$$G(\chi, \lambda_b)=\overline{\chi}(b)G(\chi), \ G(\overline{\chi})=\chi(-1)\overline{G(\chi)},$$
\ \ \ \ \ \ \ \ where $\overline{\chi}$ is the conjugate character of $\chi$ and
$$G(\chi)=G(\chi, \lambda_1)=\sum_{x\in\mathbb{F}_q^{\ast}}\chi(x)\zeta_p^{T(x)}.$$

(3) If $\chi\neq1$, then $$|G(\chi)|=\sqrt{q}.$$
\noindent From Lemma 2.1(3) we know that $G(\chi)=\sqrt{q}e^{i\theta_\chi} (0\leq\theta_\chi<2\pi)$ if $\chi\neq1.$
The exact value of Gauss sums $G(\chi)$ has been calculated for several particular cases. The following two cases will
be used in this paper.

 \textsl {Lemma 2.2:} (quadratic case, [3, Theorem 11.5.4]) Suppose that $q=p^l, p\geq3, l\geq1,$ $\eta$ is the quadratic
multiplicative character of $\mathbb{F}_q$. Then

\begin{displaymath}
 G_q(\eta)
 = \left\{ \begin{array}{ll}
(-1)^{l-1}\sqrt{q}, & \textrm{if $p\equiv1$ (mod 4)  }\\
(-1)^{l-1}(\sqrt{-1})^l\sqrt{q}, & \textrm{if $p\equiv3$ (mod 4). }
\end{array} \right.
\end{displaymath}

\textsl {Lemma 2.3:} (semiprimitive case, [3, Theorem 11.6.3]) Let $p$ be a prime number, $e\geq3$. Suppose that there
exists a positive integer $t$ such that $p^t\equiv-1$ (mod $e$). Let $t$ be the smallest positive integer satisfying
$p^t\equiv-1$ (mod $e$)(so that the multiplicative order of $p$ in $Z_e^{\ast}$ is $2t$). For $l=2ts$ $(s\geq1),$ $q=p^l$
and a multiplicative character $\chi$ of $\mathbb{F}_q$ with order $e$, we have
\begin{displaymath}
 G_q(\chi)
 = \left\{ \begin{array}{ll}
\sqrt{q}(-1)^{s-1}, & \textrm{if $p=2$  }\\
\sqrt{q}(-1)^{s-1+\frac{(p^t+1)s}{e}}, & \textrm{if $p\geq3$  }.
\end{array} \right.
\end{displaymath}

\subsection{Bounds  and  duality of  GHWs}

\ \ \ \ Several bounds on GHWs of linear codes have been given in [20]. We list three of them.

\textsl {Lemma 2.4:} Let $C$ be a linear code with parameters $[n, k]_q$. For $1\leq r \leq k$,
\begin{itemize}
\item[(1)] (Singleton type bound, [20, Corollary 3.1])
\begin{equation}
r\leq d_r(C)\leq n-k+r
\end{equation}
$C$ is called $r$-MDS code if $d_r(C)=n-k+r$.
\item[(2)] (Plotkin like bound, [20, Theorem 3.1])
\begin{equation}
d_r(C)\leq\left[\frac{n(q^r-1)q^{k-r}}{q^k-1}\right].
\end{equation}
\item[(3)](Griesmer like bound, [20, Corollary 3.3])
\begin{equation}
d_r(C)\geq \sum_{i=0}^{r-1}\left\lceil\frac{d_1(C)}{q^i}\right\rceil.
\end{equation}
\end{itemize}

The following beautiful result on dual relationship of GHWs has
been given by Wei [23].

\textsl {Lemma 2.5:} Let $C$ be a linear $[n, k]_q$ code, $k\geq 1$,
$d_r(C)\ (1\leq r\leq k)$ and $d_s(C^{\perp})\ (1\leq s \leq n-k)$
be the general Hamming weights of $C$ and its dual code $C^{\perp}$
respectively. Then
\begin{itemize}
\item[(1)] $1\leq d_1(C)<d_2(C)<\cdots<d_{k}(C)\leq n$,
\item[(2)]  $\{d_r(C): 1\leq r\leq k\}=\{1, 2, \ldots, n\}\backslash \{n+1-d_s(C^{\perp}): 1\leq s\leq n-k\}.$
\end{itemize}

\section{General Results}

\ \ \ \ From now on we fix the following notations.
\begin{itemize}
\item[(A)] $q=p^l$, where $p$ is a prime number and $l\geq 1$.
\item[(B)] $n\geq 1, (n,q)=1$, $k$ is the order of $q$ modulo $n$. Namely, $k$ is the least positive integer such that $q^k\equiv 1$ (mod $n$).
\item[(C)] $Q=q^k, Q-1=en, \theta$ is a primitive element of $\mathbb{F}_Q$. Namely, $\mathbb{F}_Q^{\ast}=\langle\theta\rangle$.
\item[(D)] $\alpha=\theta^e$ so that $\langle\alpha\rangle$ is the subgroup of $\mathbb{F}_Q^{\ast}$ with order $n$ and $\mathbb{F}_q(\alpha)=\mathbb{F}_Q.$
\item[(E)] For each integer $m\geq 1$, $\zeta_m=e^{\frac{2\pi\sqrt{-1}}{m}}.$
\item[(F)] $T_q^Q$ is the trace mapping from $\mathbb{F_Q}$ to $\mathbb{F}_q.$
\item[(G)] For a vector space $V$ over $\mathbb{F}_q$, $\begin{bmatrix}V\\r
\end{bmatrix}_q$ denotes the set of all $r$-dimentional $\mathbb{F}_q$-subspaces of $V$.
\end{itemize}

For a monic polynomial $f(x)$ in $\mathbb{F}_q[x],$ $f(0)\neq 0$,
the reciprocal polynomial of $f(x)$ is the monic polynomial
$\widehat{f}(x)=f(0)^{-1}x^df(x^{-1})\in\mathbb{F}_q[x].$ Let $h(x)$
be the minimum polynomial of $\alpha^{-1}$ over $\mathbb{F}_q$. Then
$h(x)$ is an irreducible polynomial in $\mathbb{F}_q[x],
\deg(h(x))=k$ and
$$x^n-1=h(x)g(x),\  g(x)\in \mathbb{F}_q[x],\  \deg g(x)=n-k.$$

In this paper we deal with the irreducible cyclic code $C=C(\alpha)$ over $\mathbb{F}_q$ with generating polynomial $g(x)$.
Namely,
$$C=(g(x))\subseteq \frac{\mathbb{F}_q[x]}{(x^n-1)}.$$

The parity-check polynomial of $C$ is the irreducible polynomial of
$h(x)$. The parameters of $C$ are $[n,k]_q$ where $n$ is the length
of codewords and $k=\dim_{\mathbb{F}_q}C$. It is well-known that
the dual code $C^{\perp}$ of $C$ is also cyclic with parameters $[n,
n-k]_q$ and the generating polynomial of $C^{\perp}$ is
$\widehat{h}(x).$

Our starting point in this paper is the following trace expression of $C=C(\alpha)$:
each codeword in $C$ can be uniquely expressed by
\begin{equation}
c(\beta)=(c_0, c_1,  \ldots, c_{n-1})=(T_q^Q(\beta), T_q^Q(\beta\alpha), \ldots, T_q^Q(\beta\alpha^{n-1}))\in\mathbb{F}_q^n \ (\beta\in\mathbb{F}_Q)
\end{equation}
\begin{equation}
C(\alpha)=\{c(\beta): \beta\in \mathbb{F}_Q\}.
\end{equation}

From uniqueness of trace expression (3.1), we know that for
$\beta_1, \ldots, \beta_r\in \mathbb{F}_Q$, the codewords
$c(\beta_1), \ldots, c(\beta_r)$ are $\mathbb{F}_q$-linear
independent if and only if $\beta_1, \ldots, \beta_r$ are
$\mathbb{F}_q$-linear independent. Therefore, for $1\leq r\leq k$,
the mapping
$$\Phi: \begin{bmatrix}\mathbb{F}_Q\\r
\end{bmatrix}_q\rightarrow \begin{bmatrix} C\\r
\end{bmatrix}_q, \ \Phi(H_r)=\{c(\beta): \beta\in H_r\}  \left(H_r\in \begin{bmatrix}\mathbb{F}_Q\\r
\end{bmatrix}_q\right)$$ is $\mathbb{F}_q$-linear one-to-one corresponding.

With above preparation, we can get the following two general formulas on generalized Hamming weight $d_r(C)$ of irreducible
cyclic codes $C=C(\alpha)$.

\textsl {Theorem 3.1:} Let $q=p^l$, $(n,q)=1$, $Q=q^k$ where $k$ is
the order of $q$ in $Z_{n}^{\ast}$, $Q-1=en$,
$\mathbb{F}_Q^{\ast}=\langle\theta\rangle, \alpha=\theta^e$ and
$e'=(e, \frac{Q-1}{q-1}).$ Let $C=C(\alpha)$ be the irreducible $[n,
k]_q$ cyclic code defined by (3.1) and (3.2). Then for each $r$,
$1\leq r\leq k$, $d_r(C)=n-N_r$ and
\begin{itemize}
\item[(1)] \begin{equation}
N_r=\frac{q^k-q^r}{eq^r}+\frac{1}{eq^r}\max\left\{\sum_{\tau=1}^{e'-1}G_Q(\varphi^\tau)\sum_{\beta\in
H\backslash\{0\}}\overline{\varphi}^{\tau}(\beta): H\in
\begin{bmatrix}\mathbb{F}_Q\\r
\end{bmatrix}_q\right\}
\end{equation}
where $\varphi$ is the multiplicative character of $\mathbb{F}_Q$
defined by $\varphi(\theta)=\zeta_{e'},$ and $G_Q(\varphi^\tau)
(1\leq \tau \leq e'-1)$ is the Gauss sums over $\mathbb{F}_Q.$
\item[(2)] \begin{equation}
 N_r=\frac{e'}{e}\max\left\{|H\bigcap\langle\theta^{e'}\rangle|: H\in \begin{bmatrix}\mathbb{F}_Q\\k-r
\end{bmatrix}_q\right\}.
\end{equation}
\end{itemize}
\textsl {Proof:} (1). For each $C_r\in \begin{bmatrix}C\\r
\end{bmatrix}_q$, let $H_r=\Phi^{-1}(C_r)$ be the corresponding subspace of $\mathbb{F}_Q$. Then
\begin{align}
d_r(C)& = \min\left\{|\textrm{Supp}(C_r)|: C_r\in \begin{bmatrix}C\\r
\end{bmatrix}_q\right\}\notag\\
                                  &= n-N_r,
\end{align}
where
\begin{equation}
N_r=\max\left\{N(C_r): C_r\in\begin{bmatrix}C\\r
\end{bmatrix}_q\right\}
\end{equation}
and
\begin{equation*}\begin{split}
N(C_r)& =\sharp \{i: 0\leq i\leq n-1,\ \textrm{for each}\ c=(c_0, c_1, \ldots, c_{n-1})\in C_r, c_i=0\}\\
                                  &= \sharp \{i: 0\leq i\leq n-1,\ \textrm{for each}\  \beta\in H_r,
                                  T_q^Q(\beta\alpha^i)=0\}.
\end{split}\end{equation*}
Let $\{\beta_1, \ldots, \beta_r\}$ be an $\mathbb{F}_q$-basis of
$H_r$. It is easy to see that for each $i$, $0\leq i\leq n-1$,
$T_q^Q(\beta\alpha^i)=0$ for all $\beta\in H_r$ if and only if
$T_q^Q(\beta_\lambda\alpha^i)=0\ (1\leq\lambda\leq r).$ Therefore

\begin{equation*}\begin{split}
N(C_r)& =\frac{1}{q^r}\sum_{i=0}^{n-1}\left(\sum_{x_1\in
\mathbb{F}_q}\zeta_p^{T_p^q(T_q^Q(\beta_1\alpha^i)x_1)}\right)\ldots
\left(\sum_{x_r\in \mathbb{F}_q}\zeta_p^{T_p^q(T_q^Q(\beta_r\alpha^i)x_r)}\right)\\
                                  &= \frac{1}{q^r}\sum_{x_1, \ldots, x_r\in \mathbb{F}_q}\sum_{i=0}^{n-1}\zeta_p^{T_p^Q\left(\alpha^i(\beta_1x_1+\cdots+\beta_rx_r)\right)}\\
                                  &= \frac{1}{q^r}\sum_{\beta\in H_r}\sum_{i=0}^{n-1}\zeta_p^{T_p^Q(\beta\alpha^i)}\\
                                  &=\frac{n}{q^r}+\frac{1}{q^r}\sum_{\beta\in H_r\backslash\{0\}}\sum_{i=0}^{n-1}\zeta_p^{T_p^Q(\beta\alpha^i)}.
\end{split}\end{equation*}

Let $\chi$ be the multiplicative character of
$\mathbb{F}_Q^{\ast}=\langle\theta\rangle$ defined by
$\chi(\theta)=\zeta_e$. From $\alpha=\theta^e$ we know that for each
$x\in\mathbb{F}_Q^{\ast}$,
\begin{displaymath}
 \sum_{\lambda=0}^{e-1}\chi^\lambda(x)
 = \left\{ \begin{array}{ll}
e, & \textrm{if $x\in \langle\alpha\rangle$}\\
0, & \textrm{otherwise}.
\end{array} \right.
\end{displaymath}
Therefore
\begin{align}
N(C_r)& =\frac{n}{q^r}+\frac{1}{eq^r}\sum_{\beta\in H_r\backslash\{0\}}\sum_{x\in\mathbb{F}_Q^{\ast}}\zeta_p^{T_p^Q(\beta x)}\sum_{\lambda=0}^{e-1}\chi^\lambda(x)\notag\\
                                  &= \frac{n}{q^r}+\frac{1}{eq^r}\sum_{\beta\in H_r\backslash\{0\}}\sum_{\lambda=0}^{e-1}
                                  \overline{\chi}^\lambda(\beta)G_Q(\chi^\lambda)\ \ \ \ \ \ \ \ \ \ \ \ \ \ \ \ \ \ \ \ \ \ \ \ \ \ \ \ \ \textrm{(by Lemma 2.1(2))}\notag\\
                                  &= \frac{n}{q^r}+\frac{1}{eq^r}\sum_{\lambda=0}^{e-1}G_Q(\chi^\lambda)\sum_{\beta\in H_r\backslash\{0\}}\overline{\chi}^\lambda(\beta).
                                  \end{align}

Since $H_r$ is $\mathbb{F}_q$-vector space, for each
$a\in\mathbb{F}_q^{\ast},$
$a(H_r\backslash\{0\})=H_r\backslash\{0\}$. If there exists
$a\in\mathbb{F}_q^{\ast}$ such that
$\overline{\chi}^\lambda(a)\neq1$, then
$$\sum_{\beta\in H_r\backslash\{0\}}\overline{\chi}^\lambda(\beta)=\sum_{\beta\in H_r\backslash\{0\}}\overline{\chi}^\lambda(\beta a)=\overline{\chi}^\lambda(a)\sum_{\beta\in H_r\backslash\{0\}}\overline{\chi}^\lambda(\beta)$$
we get $\sum_{\beta\in H_r\backslash\{0\}}\overline{\chi}^\lambda(\beta)=0$. Thus the formula (3.7) becomes
$$N(C_r)=\frac{n}{q^r}+\frac{1}{eq^r}\sum^{e-1}_{\lambda=0\atop \chi^\lambda(\mathbb{F}_q^{\ast})=1 }G_Q(\chi^\lambda)\sum_{\beta\in H_r\backslash\{0\}}\overline{\chi}^\lambda(\beta).$$

From
$\mathbb{F}_q^{\ast}=\left\langle\theta^{\frac{Q-1}{q-1}}\right\rangle$
and $\chi(\theta)=\zeta_e$ we know that
\begin{equation*}\begin{split}
\chi^\lambda(\mathbb{F}_q^{\ast})=1& \Longleftrightarrow 1=\chi^\lambda(\theta^{\frac{Q-1}{q-1}})=\zeta_e^{\lambda\cdot\frac{Q-1}{q-1}}\Longleftrightarrow e\bigg|\lambda\cdot\frac{Q-1}{q-1}\\
                                  &\Longleftrightarrow \frac{e}{e'}\bigg|\lambda \ \left(e'=\left(e, \frac{Q-1}{q-1} \right)\right)\ .
                                  \end{split}\end{equation*}
Let $e=e'd$. By $\varphi(\theta)=\zeta_{e'}$ and $\chi(\theta)=\zeta_e$ we know that $\varphi=\chi^d$. Therefore

\begin{equation*}\begin{split}
\{\chi^\lambda: 0\leq \lambda \leq e-1, \chi^\lambda(\mathbb{F}_q^{\ast})=1\}=& \{\chi^{d\tau}: 0\leq \tau \leq e'-1\}\\
                                  =&\{\varphi^{\tau}: 0\leq \tau \leq e'-1\}.
                                  \end{split}\end{equation*}

And then
\begin{align}
N(C_r)& =\frac{n}{q^r}+\frac{1}{eq^r}\sum_{\tau=0}^{e'-1}G_Q(\varphi^\tau)\sum_{\beta\in H_r\backslash\{0\}}\overline{\varphi}^\tau(\beta)\notag\\
                                  &= \frac{n}{q^r}+\frac{1}{eq^r}\left[-\left(|H_r|-1\right)+\sum_{\tau=1}^{e'-1}G_Q(\varphi^\tau)\sum_{\beta\in H_r\backslash\{0\}}\overline{\varphi}^\tau(\beta)\right]\notag\\
                                  &= \frac{q^k-1}{eq^r}-\frac{q^r-1}{eq^r}+\frac{1}{eq^r}\sum_{\tau=1}^{e'-1}G_Q(\varphi^\tau)\sum_{\beta\in H_r\backslash\{0\}}\overline{\varphi}^\tau(\beta).
                                  \end{align}
The formula (3.3) is derived from (3.5), (3.6) and (3.8).

(2). It is well-known that for $x\in\mathbb{F}_Q$,
\begin{displaymath}
 \sum_{\gamma\in H_r^{\perp}}\zeta_p^{T_p^Q(\gamma x)}
 = \left\{ \begin{array}{ll}
|H_r^{\perp}|=q^{k-r}, & \textrm{if $x\in H_r$}\\
0, & \textrm{otherwise}
\end{array} \right.
\end{displaymath}
where $H_r^{\perp}$ is the dual of $H_r$ in $C.$ Therefore for $\varphi^\tau\neq 1$ (we assume $\varphi^\tau (0)=0),$
\begin{equation*}\begin{split}
\sum_{\beta\in H_r\backslash\{0\}}\overline{\varphi}^\tau(\beta)& =\sum_{\beta\in H_r}\overline{\varphi}^\tau(\beta)=\frac{1}{q^{k-r}}\sum_{\beta\in\mathbb{F}_Q}\sum_{\gamma\in H_r^{\perp}}\zeta_p^{T_p^Q(\beta\gamma)}\overline{\varphi}^\tau(\beta)\\
                                  &= \frac{1}{q^{k-r}}\sum_{\gamma\in H_r^{\perp}\backslash\{0\}}\sum_{\beta\in\mathbb{F}_Q^{\ast}}\zeta_p^{T_p^Q(\beta\gamma)}\overline{\varphi}^\tau(\beta)\\
                                  &= \frac{1}{q^{k-r}}\sum_{\gamma\in H_r^{\perp}\backslash\{0\}}G_Q(\overline{\varphi}^\tau){\varphi}^\tau(\gamma).
                                  \end{split}\end{equation*}
Then (3.8) becomes
\begin{equation*}\begin{split}
N(C_r)& =\frac{q^k-q^r}{eq^r}+\frac{1}{eq^k}\sum_{\tau=1}^{e'-1}G_Q(\varphi^\tau)G_Q(\overline{\varphi}^\tau)\sum_{\gamma\in H_r^{\perp}\backslash\{0\}}\varphi^{\tau}(\gamma)\\
                                  &= \frac{q^k-q^r}{eq^r}+\frac{1}{e}\sum_{\tau=1}^{e'-1}\varphi^{\tau}(-1)\sum_{\gamma\in H_r^{\perp}\backslash\{0\}}\varphi^\tau(\gamma)\ \ \ \textrm{(by Lemma 2.1(2))}\\
                                  &= \frac{q^k-q^r}{eq^r}+\frac{1}{e}\sum_{\gamma\in H_r^{\perp}\backslash\{0\}}
                                  \sum_{\tau=1}^{e'-1}\varphi^\tau(\gamma)\ \ \ \ \ \ \ \ \ \textrm{(since $-H_r^{\perp}=H_r^{\perp}$})\\
                                  &=\frac{q^k-q^r}{eq^r}-\frac{|H_r^{\perp}|-1}{e}+\frac{1}{e}\sum_{\gamma\in H_r^{\perp}\backslash\{0\}}
                                  \sum_{\tau=0}^{e'-1}\varphi^\tau(\gamma)\\
                                  &=\frac{q^k-q^r}{eq^r}-\frac{q^{k-r}-1}{e}+\frac{e'}{e}|H_r^{\perp}\bigcap\langle\theta^{e'}\rangle|=\frac{e'}{e}|H_r^{\perp}\bigcap\langle\theta^{e'}\rangle|.
                                  \end{split}\end{equation*}
Then we get (3.4) from (3.6). This completes the proof of Theorem 3.1. \qed

As a direct consequence of Theorem 3.1, we consider the case $e'=1$.
In this case, either by (3.3), where the summation in right-hand
side is zero, or by (3.4) where $H_r^{\perp}\backslash\{0\}\subseteq
\mathbb{F}_Q^{\ast}=\langle\theta\rangle$ and
$|H_r^{\perp}\bigcap\langle\theta\rangle|=|H_r^{\perp}\backslash\{0\}|=q^{k-r}-1$,
we get the following simple result.

\textsl {Corollary 3.2:} (case $e'=1$) Let $q=p^l, Q=q^k, Q-1=en,$
and $C$ be the irreducible cyclic code in Theorem 3.1. If $e'=(e,
\frac{Q-1}{q-1})=1$ (which means that $e\mid q-1$ and $(e, k)=1),$
then
$$d_r(C)=\frac{q^k-q^{k-r}}{e}\ \ \ \ (0\leq r\leq k).$$

\textsl {Remark:} (1) From $d_r(C)=\frac{n(q^r-1)q^{k-r}}{q^k-1}$ we know that in the case $e'=1$, $d_r(C)$ meets  the Plotkin type bound
and the Griesmer type bound for all $r$, $1\leq r\leq k$. Namely, both of (2.2) and (2.3) in Lemma 2.4 are equality.

(2) From Lemma 2.5(2) we know that for case $e'=1$ and $k\geq 2$,
\begin{eqnarray*}
&&\{d_1(C^{\perp}), \ldots, d_{n-k}(C^{\perp})\}\\
&&=\{1, 2, \ldots, n\}\backslash\{n+1-d_k(C), \ldots, n+1-d_1(C)\}\\
&&=\{1, 2, \ldots, n\}\backslash\left\{n+1-n, n+1-\left(n-\frac{q^k-q^{k-1}}{eq^{k-1}}\right),\ldots, n+1-\left(n-\frac{q^k-q}{eq}\right)\right\} \\
&&= \{1, 2, \ldots, n\}\backslash\left\{1, 1+\frac{q-1}{e}, \ldots,
1+\frac{q^{k-1}-1}{e}\right\}.
\end{eqnarray*}
Therefore
$$d_{n-k-i}(C^{\perp})=n-i\ \ \ \ \ \left(0\leq i\leq n-\left(\frac{q^k-1}{e}+2\right)\right).$$
Namely, $d_r(C^{\perp})=n-(n-k)+r$ for $\frac{q^{k-1}-1}{e}+2-k\leq
r\leq n-k.$ By (2.1) in Lemma 2.4 we know that $C^{\perp}$ is
$r$-MDS code for $r$ satisfying $\frac{q^{k-1}-1}{e}+2-k\leq r\leq
n-k$.

From now on we assume $e'\geq 2$. This can be devided two subcases:
\begin{itemize}
\item[(A)] $e'|q-1$. In this case, $e'=(e, \frac{q^k-1}{q-1})|\frac{q^k-1}{q-1}$, we get $e'|k.$
\item[(B)] $e'\nmid q-1.$
\end{itemize}

For case $(A)$ we can determine $d_r(C)$ for larger $r$.

\textsl {Theorem 3.3:} Suppose that $e'=(e, \frac{Q-1}{q-1})|q-1$
and $e'\geq 2$. $C$ is the irreducible cyclic code in Theorem
3.1 with parameters $[n, k]_q$. Let $k=e'm$. Then for $k-m\leq r\leq
k$,
$$d_r(C)=n-\frac{e'}{e}(q^{k-r}-1).$$

\textsl {Proof:} From $e'\mid q-1$ we get
$$\left(e', \frac{q^k-1}{q^m-1}\right)=(e', q^{(e'-1)m}+q^{(e'-2)m}+\cdots+q^m+1)=(e', e')=e'.$$
Namely, $e'|\frac{q^k-1}{q^m-1}$. Consider the subfield
$\mathbb{F}_{q^m}$ of $\mathbb{F}_{q^k}$. From
$\mathbb{F}_{q^m}^{\ast}=\langle\theta^{\frac{q^k-1}{q^m-1}}\rangle$
we get $\mathbb{F}_{q^m}^{\ast}\subseteq\langle\theta^{e'}\rangle$.
For $k-m\leq r\leq k$, we have $0\leq k-r\leq m$ so that we can take
$H$ as a $(k-r)$-dimensional $\mathbb{F}_q$-subspace of
$\mathbb{F}_{q^m}$. Then
$|H\bigcap\langle\theta^{e'}\rangle|=|H\backslash\{0\}|=q^{k-r}-1$
reachs the maximum value in the right-hand side of (3.4).
Therefore $N_r=\frac{e'}{e}(q^{k-r}-1)$ and
$d_r(C)=n-N_r=n-\frac{e'}{e}(q^{k-r}-1).$  \qed

For case (B), the following result can be proved in similar way.

\textsl {Theorem 3.4:} Let $C$ be the irreducible cyclic code with parameter $[n, k]_q$. Suppose that $e'=(e, \frac{Q-1}{q-1})\nmid q-1$. Let $m$ be a positive factor of $k$ such that $e'|\frac{q^k-1}{q^m-1}.$ Then for $k-m\leq r\leq k$,
$d_r(C)=n-\frac{e'}{e}(q^{k-r}-1).$

Theorems 3.3 and 3.4 are general results for case $e'\geq2$ by using
formula (3.4). On the other hand, formula (3.3) involves related
Gauss sums $G_Q(\varphi^\tau) (1\leq\tau\leq e'-1)$. In next section
we will get further results on $d_r(C)$ for smaller $r$ by using
(3.3) in several particular cases where the Gauss sums
$G_Q(\varphi^\tau) (1\leq\tau\leq e'-1)$ have the same value.

\section{Further Results on $d_r(C)$}

In Section 3 we have determined the weight hierarchy of the irreducible cyclic code $C$ in case $e'=1$ and the values of $d_r(C)$
for larger $r$ in case $e'\geq 2$ by (3.4). In this section we determine $d_r(C)$ for smaller $r$ by using formula (3.3) for case
$e'=2$ and semiprimitive case.

(I) $e'=2$ case

\textsl {Theorem 4.1:} Let $q=p^l, Q=q^k, Q-1=en$ and $C$ be the irreducible cyclic code with parameters $[n, k]_q$ given in
Theorem 3.1. Suppose that $e'=(e, \frac{Q-1}{q-1})=2$ so that $p$ is an odd prime number and $k=2s$ is even. Then the weight hierarchy of
$C$ is
\begin{displaymath}
 d_r(C)
 = \left\{ \begin{array}{ll}
\frac{1}{e}(q^s-1)(q^{s-r}+1), & \textrm{for $0\leq r\leq s$}\\
\frac{1}{e}(q^{2s}-2q^{2s-r}+1), & \textrm{for $s\leq r\leq 2s=k$}.
\end{array} \right.
\end{displaymath}

\textsl {Proof:} Let $\eta$ be the quadratic multiplicative character of $\mathbb{F}_Q$. Namely, $\eta(\langle\theta^2\rangle)=1$ and
$\eta(\theta\langle\theta^2\rangle)=-1$. We have $d_r(C)=n-N_r$ and formula (3.3) becomes
\begin{equation}
N_r=\frac{q^{2s}-q^r}{eq^r}+\frac{1}{eq^r}\max\left\{G_Q(\eta)\sum_{\beta\in
H\backslash\{0\}}\eta(\beta): H\in \begin{bmatrix}\mathbb{F}_Q\\r
\end{bmatrix}_q\right\}.
\end{equation}
By Lemma 2.2, $G_Q(\eta)=\varepsilon\sqrt{Q}=\varepsilon q^s$ where
$\varepsilon=1$ or $-1.$ Now we consider the subfield
$\mathbb{F}_{q^s}$ of $\mathbb{F}_{q^{2s}}=\mathbb{F}_Q$. From
$\mathbb{F}_{q^s}^{\ast}=\langle\theta^{\frac{q^{2s}-1}{q^{s}-1}}\rangle=\langle\theta^{q^{s}+1}\rangle\subseteq\langle\theta^2\rangle$
we get $\eta(\mathbb{F}_{q^s}^{\ast})=1$. For $0\leq r\leq s$, if
$\varepsilon=1$, we take any $r$-dimensional $\mathbb{F}_q$-subspace
$H$ of $\mathbb{F}_{q^s}.$ We have $G_Q(\eta)\sum_{\beta\in
H\backslash\{0\}}\eta(\beta)=\sqrt{Q}\sum_{\beta\in
H\backslash\{0\}}\eta(\beta)=\sqrt{Q}|H\backslash\{0\}|=(q^r-1)q^s.$ If
$\varepsilon=-1,$ we take $H'=\theta  H
\in\begin{bmatrix}\mathbb{F}_Q\\r
\end{bmatrix}_q. $ Since $\eta(H')=-1$ we have $G_Q(\eta)\sum_{\beta\in H'\backslash \{0\}}\eta(\beta)=-\sqrt{Q}(-(q^r-1))=(q^r-1)q^s$. Therefore the maximal value in right-hand side of (4.1) is $(q^r-1)q^s$  so that
$$d_r(C)=\frac{q^{2s}-q^r}{eq^r}+\frac{(q^r-1)q^s}{eq^r}=\frac{1}{e}(q^s-1)(q^{s-r}+1)\ \  (\textrm{for} \ 0\leq r\leq s).$$

On the other hand, $e'=2|\frac{q^{2s}-1}{q^s-1}$. By Theorem 3.4 we get
$$d_r(C)=n-\frac{2}{e}(q^{k-r}-1)=\frac{1}{e}(q^{2s}-2q^{2s-r}+1)\ (\textrm{for} \ s\leq r\leq 2s=k).$$

(II) Semiprimitive Case

From now on we assume that $e'\geq 3$. The semiprimitive (or called ``self-conjugated") case means that the following condition $(\ast)$ is satisfied.

$(\ast)$ There exists integer $t$ such that $p^t\equiv -1 \textrm {(mod}\ e')$.

 We always assume that $t$  is the smallest positive integer satisfying $p^t\equiv -1 \textrm {(mod}\ e')$. Then the order of $p$ in $Z_{e'}^{\ast}$ is
$2t$ and for $q=p^l, Q=q^k=p^{lk},$ from $e'|Q-1$ we get $2t|lk$. Namely, $lk=2ts \ (s\in Z$).

As in Section 3, for $e'\geq 3$ we divide the following two
subcases:
\begin{itemize}
\item[(A)] $e'|q-1$, namely $2t|l$. In this case $e'|k$.
\item[(B)] $e'\nmid q-1$, namely $2t\nmid l$.
\end{itemize}

Theorem 3.1 shows that $d_r(C)=n-N_r$ and $N_r$ has the expression
(3.3). Namely,
\begin{equation}
N_r=\frac{q^k-q^r}{eq^r}+\frac{1}{eq^r}\max\left\{\sum_{\tau=1}^{e'-1}G_Q(\varphi^{\tau})\sum_{\beta\in
H\backslash\{0\}}\varphi^{\tau}(\beta): H\in
\begin{bmatrix}\mathbb{F}_Q\\r
\end{bmatrix}_q\right\}
\end{equation}

 \noindent where $\varphi$ is the multiplicative character of $\mathbb{F}_Q^{\ast}=\langle\theta\rangle$ with order $e'$ defined by $\varphi(\theta)=\zeta_{e'}$.

For each $\tau$, $1\leq \tau\leq e'-1$, the order of $\varphi^\tau$ is $e_\tau=\frac{e'}{(e', \tau)}$. From the semiprimitive condition $(\ast)$ we have
$p^t\equiv -1 \textrm {(mod}\ e_{\tau})$. Let $t_\tau$ be the least positive integer satisfying $p^{t_{\tau}}\equiv -1 \textrm {(mod}\ e_{\tau})$. Then the
order of $p$ in $Z_{e^{\tau}}^{\ast}$ is $2t_\tau$ so that $2t_\tau|2t$, and $t=t_\tau m_\tau\ (m_\tau\in Z).$ Then $lk=2ts=2t_{\tau}m_{\tau}s$. By Lemma 2.3, for
$1\leq \tau \leq e'-1$ we have
$$G_Q(\varphi^{\tau})=\sqrt{Q}\varepsilon_{\tau},$$
where
 \begin{align}
 \varepsilon_{\tau}
 = \left\{ \begin{array}{ll}
(-1)^{m_{\tau}s-1}, & \textrm{for $p=2$} \\
(-1)^{m_{\tau}s-1+\frac{(p^{t_{\tau}}+1)m_{\tau}s}{e_{\tau}}}, & \textrm{for $p\geq 3$}.
\end{array} \right.
\end{align}

Firstly we consider the subcase $(A)$.

\textsl {Theorem 4.2:} Let $q=p^l, Q=q^k, Q-1=en, e'=(e, \frac{Q-1}{q-1})\geq3$. Assume that the semiprimitive condition $(\ast)$ holds where
$t$ is the least positive integer satisfying $p^t\equiv -1 \textrm {(mod}\ e'),$ so that $lk=2ts$. Let $C$ be the irreducible cyclic code in Theorem 3.1
with parameters $[n, k]_q$. If $e'|q-1$ (which means $2t|l$) so that $k=k'e'\ (k'\in Z).$ Then
\begin{itemize}
\item[(1)] For $k-k'\leq r\leq k$,  $d_r(C)=n-\frac{e'}{e}(q^{k-r}-1).$
\item[(2)] If $s$ is even, then for $1\leq r\leq k'$,
$$d_r(C)=n-\frac{1}{eq^r}(q^k-q^r+q^{k/2}(q^r-1)).$$
\item[(3)] If $2\nmid se'$, then for $1\leq r\leq k'$,
$$d_r(C)=n-\frac{1}{eq^r}(q^k-q^r+(e'-1)q^{k/2}(q^r-1)).$$
\end{itemize}

\textsl {Proof:} (1). For $k-k'\leq r\leq k$ we have $0\leq k-r\leq
k'$. Let $Q'=q^{k'}$. From $e'\mid q-1\mid Q'-1$ we know that $(e',
\frac{Q-1}{Q'-1})=(e', \frac{k}{k'})=(e', e')=e'.$ Therefore
$e'|\frac{Q-1}{Q'-1}$ and the conclusion of (1) can be derived from
Theorem 3.4.

(2). If $2|s$, then all $G(\varphi^{\tau}) \ (1\leq \tau \leq e'-1)$ are $-\sqrt{Q}$ by formula (4.3). Then formula (4.2) becomes
\begin{align}
N_r& =\frac{q^k-q^r}{eq^r}-\frac{q^{k/2}}{eq^r}\min\left\{\sum_{\beta\in H\backslash\{0\}}\sum_{\tau=1}^{e'-1}\varphi^{\tau}(\beta): H\in\begin{bmatrix}\mathbb{F}_Q\\r\end{bmatrix}_q\right\}\notag\\
                                  &= \frac{q^k-q^r}{eq^r}+ \frac{q^{k/2}(q^r-1)}{eq^r}-\frac{q^{k/2}}{eq^r}\min\left\{\sum_{\beta\in H\backslash\{0\}}\sum_{\tau=0}^{e'-1}\varphi^{\tau}(\beta): H\in\begin{bmatrix}\mathbb{F}_Q\\r\end{bmatrix}_q\right\}\notag\\
                                   &=\frac{q^k-q^r+q^{k/2}(q^r-1)}{eq^r}-\frac{e'q^{k/2}}{eq^r}\min\left\{|H\bigcap\langle\theta^{e'}\rangle|: H\in\begin{bmatrix}\mathbb{F}_Q\\r\end{bmatrix}_q\right\}.
                                  \end{align}
For $1\leq r\leq k'$, we take an $r$-dimensional $\mathbb{F}_q$-subspace $H'$ of $\mathbb{F}_{Q'}$. Then for $\mathbb{F}_q$-subspace $H=\theta H'$ of $\mathbb{F}_Q$ we get the minimum value
$|H\bigcap\langle\theta^{e'}\rangle|=0$ in the right-hand side of (4.4). Therefore
$$d_r(C)=n-N_r=n-\frac{q^k-q^r+q^{k/2}(q^r-1)}{eq^r}.$$

(3) If $2\nmid se'$, then $s$, all $e_\tau (|e') (1\leq \tau \leq e'-1)$ are odd. From $p^t\equiv -1$ (mod $e'$) and $p^{t_{\tau}}\equiv -1$(mod $e_{\tau}$), we know that all $m_{\tau}=t/t_{\tau} (1\leq \tau\leq e'-1)$ are odd. By (4.3) we get $\varepsilon_\tau=1$ and $G_Q(\varphi^{\tau})=q^{k/2}$ for all $\tau$, $1\leq\tau\leq e'-1$.
 Then formula (4.2) becomes
\begin{align}
N_r& =\frac{q^k-q^r}{eq^r}+\frac{q^{k/2}}{eq^r}\max\left\{\sum_{\beta\in H\backslash\{0\}}\sum_{\tau=1}^{e'-1}\varphi^{\tau}(\beta): H\in\begin{bmatrix}\mathbb{F}_Q\\r\end{bmatrix}_q\right\}\notag\\
                                  &= \frac{q^k-q^r}{eq^r}- \frac{q^{k/2}(q^r-1)}{eq^r}+\frac{q^{k/2}}{eq^r}\max\left\{\sum_{\beta\in H\backslash\{0\}}\sum_{\tau=0}^{e'-1}\varphi^{\tau}(\beta): H\in\begin{bmatrix}\mathbb{F}_Q\\r\end{bmatrix}_q\right\}\notag\\
                                   &=\frac{q^k-q^r-q^{k/2}(q^r-1)}{eq^r}+\frac{e'q^{k/2}}{eq^r}\max\left\{|H\bigcap\langle\theta^{e'}\rangle|: H\in\begin{bmatrix}\mathbb{F}_Q\\r\end{bmatrix}_q\right\}.
                                  \end{align}
For $1\leq r\leq k'$, we take $H$ as a $r$-dimensional $\mathbb{F}_q$-subspace of $\mathbb{F}_{Q'}$. Then $|H\bigcap\langle\theta^{e'}\rangle|=q^r-1$ is the maximal value in the right-hand
side of (4.5). Therefore
\begin{equation*}\begin{split}
d_r(C)& =n-N_r=n-\frac{1}{eq^r}\left[q^k-q^r-q^{k/2}(q^r-1)+e'q^{k/2}(q^r-1)\right]\\
                                  &=n-\frac{1}{eq^r}\left[q^k-q^r+(e'-1)q^{k/2}(q^r-1)\right]. \qed
                                  \end{split}\end{equation*}

Now we consider the subcase $(B)$.

\textsl {Theorem 4.3:} Let $q=p^l, Q=q^k, Q-1=en$ and $e'=(e, \frac{Q-1}{q-1})\geq 3.$
Suppose that the semiprimitive condition $(\ast)$ holds where $t$ is the least positive number satisfying $p^t\equiv -1$ (mod $e'$), so that
$lk=2ts$. Let $C$ be the irreducible cyclic code in Theorem 3.1 with parameters $[n, k]_q$. Suppose that $e'\nmid q-1$ which means that $2t \nmid l$. Let
\begin{equation}
l=2^al',\ k=2^bk',\ t=2^{c}t',\ 2\nmid l'k't'
\end{equation}
and assume that $c\geq a$. Let $m=2^{c-a}k'$. Then
\begin{itemize}
\item[(1)] For $k-m\leq r\leq k$, $d_r(C)=n-\frac{e'}{e}(q^{k-r}-1).$
\item[(2)] If $2|s$, then for $1\leq r\leq m$,
$$d_r(C)=n-\frac{1}{eq^r}(q^k-q^r+q^{k/2}(q^r-1)).$$
\item[(3)] If $2\nmid se'$, then for $1\leq r\leq m$,
$$d_r(C)=n-\frac{1}{eq^r}(q^k-q^r+(e'-1)q^{k/2}(q^r-1)).$$
\end{itemize}

\textsl {Proof:} (1). Consider the finite field $\mathbb{F}_{Q'}$ where $Q'=q^m=p^{lm}.$ From (4.6) and $lk=2ts$ we know that
$$lm=l\cdot2^{c-a}k'=2^cl'k'=2^ct's'=ts'$$
where $s'$ is the odd part of $s$. From $lm=ts'|2ts=lk$ we know that $\mathbb{F}_q\subseteq\mathbb{F}_{\mathbb{Q}'}\subseteq\mathbb{F}_{\mathbb{Q}}$.
Since $s'$ is odd, we have
\begin{equation}
Q'=p^{ts'}\equiv -1 \ (\textrm{mod}\ e').
\end{equation}
Moreover, from $lk=2ts$ and (4.6) we get $a+b\geq c+1$.
Let $\tau=\frac{k}{m}=\frac{2^bk'}{2^{c-a}k'}=2^{a+b-c},$ then $2|\tau$ and by (4.7),
$$\frac{Q-1}{Q'-1}=\frac{p^{lk}-1}{p^{lm}-1}=\sum_{\lambda=0}^{\tau-1}p^{lm\lambda}=\sum_{\lambda=0}^{\tau-1}p^{ts'\lambda}
\equiv\sum_{\lambda=0}^{\tau-1}(-1)^{\lambda}\equiv0\ (\textrm{mod}\
e')$$ which means that
$\mathbb{F}_{Q'}^{\ast}=\left\langle\theta^{\frac{Q-1}{Q'-1}}\right\rangle\subseteq\langle\theta^{e'}\rangle.$
Then the conclusion of (1) can be derived from Theorem 3.4.

(2) and (3) can be proved by the same way as Theorem 4.2(2) and (3).
\qed

In fact, under the condition $2\nmid se'$, we can determine the total weight hierarchy of $C$.

\textsl {Corollary 4.4:} Under the assumptions $c\geq a$ and $2\nmid se'$, the weight hierarchy of $C$ is
\begin{displaymath}
 d_r(C)
 = \left\{ \begin{array}{ll}
n-\frac{1}{eq^r}(q^k-q^r+(e'-1)q^{k/2}(q^r-1)), & \textrm{for $0\leq r\leq k/2$}\\
n-\frac{e'}{e}(q^{k-r}-1), & \textrm{for $\frac{k}{2}\leq r\leq k$}.
\end{array} \right.
\end{displaymath}

\textsl {Proof:} In this case, $b=c-a+1$ and
$m=2^{c-a}k'=2^{b-1}k'=k/2$. The conclusion can be derived from (1)
and (3) of Theorem 4.3. \qed

\textsl {Remark:} In binary case $(p=2)$, $e'$ is always odd. The assumption of Corollary 4.4 becomes to be $c\geq a$ and $2\nmid s$.
Particularly for $q=2, l=1$ so that $a=0$, the assumption becomes to be $2\nmid s,$ since $c\geq 0(=a)$ is always true.
This is just the Theorem 3 of [14], but two proofs are quite different.

\section{Conclusion}

\ \ \ \ By using Gauss sums, we present two formulas (3.3) and (3.4) on generalized Hamming weight $d_r(C)$ for irreducible cyclic
code $C$. For cases $e'=1$ and 2, the weight hierarchy $\{d_r(C): 1\leq r\leq k\}$ has been completely determined, so that
the weight hierarchy of the dual code $C^{\perp}$ can also be determined by Lemma 2.5(2). In general case we can determine
$d_r(C)$ for smaller $r$ by (3.3) and bigger $r$ by (3.4). For intermediate values of $r$, formulas (3.3) or (3.4) involve the
extreme value of character sums $\sum_{x\in H\backslash \{0\}}\varphi^{\tau}(x)$ or $|H\bigcap\langle\theta^{e'}\rangle|$
where $H$ pass through all $r$-dimensional or $(k-r)$-dimensional $\mathbb{F}_q$-subspace of $\mathbb{F}_{q^k}$ which may have
their own interests in finite field theory.

The generalized Hamming weights of reducible cyclic codes will be the next topic for further research. Helleseth and Kumar
[13] have determined the weight hierarchy for binary Kasami codes which are cyclic codes $C$ over $\mathbb{F}_{q^2} (q=2^m)$
such that $C^{\perp}$ has two roots $\theta$ and $\theta^{q+1}$ where $\mathbb{F}_{q^2}^{\ast}=\langle\theta\rangle$.
Recently many results on weight distribution of reducible cyclic codes for ordinary Hamming weight $d(C)=d_1(C)$ have been developed with various
techniques. We hope that such techniques may help to determine generalized Hamming weight $d_r(C)$ for $r\geq 2$.

\end{document}